# Arbitrary manipulation of nonlinear optical processes

Jian Zheng and Masayuki Katsuragawa[*]

Department of Engineering Science, University of Electro-Communications,
1-5-1, Chofugaoka, Chofu, Tokyo 182-8585, Japan
[*]*katsuragawa@uec.ac.jp*

**Abstract**

Nonlinear optical processes are governed by the relative-phase relationships among the relevant electromagnetic fields in these processes. In this Letter, we describe the physics of arbitrary manipulation of nonlinear optical processes (AMNOP) by artificial control of relative phases. As a typical example, we show freely designable optical-frequency conversions to extreme spectral regions, mid-infrared and vacuum-ultraviolet, with near-unity quantum efficiencies. Furthermore, we show that such optical-frequency conversions can be realized by using a surprisingly simple technology where transparent plates are placed in a nonlinear optical medium and their positions and thicknesses are adjusted precisely. In a numerical experiment assuming practically applicable parameters in detail, we demonstrate a single-frequency tunable laser that covers the whole vacuum-ultraviolet spectral range of 120 to 200 nm.

Nonlinear optical processes are dominated by the relative-phase relationships among the relevant electromagnetic fields in the processes. The representative example could be phase-matching [1]. Since the birth of nonlinear optics in 1961 [1, 2], huge efforts have been devoted to studying how we can achieve phase-matching to realize efficient nonlinear optical phenomena. Various technologies have been developed with great success [3, 4, 5], including those that use crystal birefringence [6, 7], the angle distributions of the relevant laser fields [8], the concept of quasi-phase-matching [1, 9, 10], or the control of optical properties by embedding metamaterial structures [11]. However, the control of relative-phase relationship is not necessarily just for achieving phase-matching. In general, there are various possibilities. Although it has barely been discussed, if we could manipulate such relative phases arbitrarily, beyond phase-matching, then we would be able to freely control nonlinear optical processes. Is such control possible in reality? In this Letter, we discuss this attractive possibility by using as an example a specific nonlinear optical process, namely a Raman-resonant four-wave-mixing process, including its high-order processes [12, 13, 14].

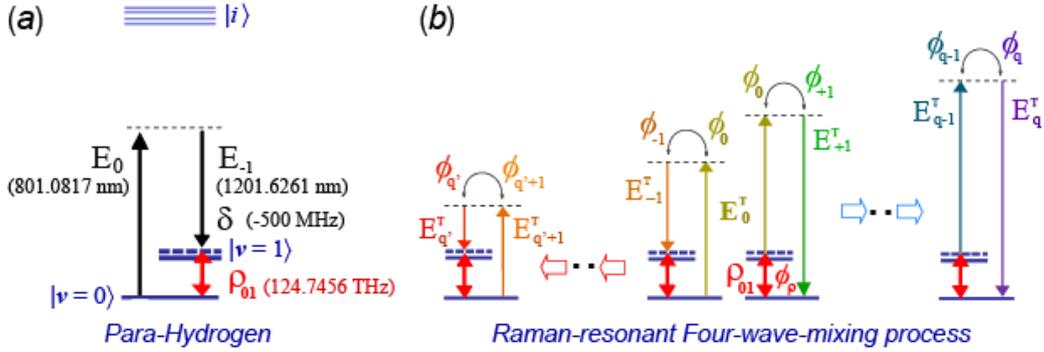

**Fig. 1. Scheme of Raman-resonant four-wave-mixing process in parahydrogen.** *a*, Adiabatic driving of vibrational coherence at a Raman transition of $v = 0$ to 1. *b*, High-order four-wave-mixing process initiated from the incident third-laser field, $E_0^T$.

Figure 1 illustrates the scheme of the Raman-resonant four-wave-mixing process. We employ gaseous parahydrogen as a nonlinear optical medium and focus on the pure vibrational Raman transition of $v = 0, J = 0$ to $v = 1, J = 0$ at 125.7451 THz [14, 15]. First, we adiabatically drive a high coherence between these two vibrational levels, which is achieved by applying two laser-fields, $E_0$ and $E_{-1}$, and controlling the small two-photon detuning, $\delta$, from the Raman resonance (Fig. 1) [12]. This adiabatic excitation process of high coherence, $\rho_{01}$, in turn deeply modulates the two driving laser fields, $E_0$ and $E_{-1}$ and generates the high-order Stokes and anti-Stokes components, $E_q$ ($q$: integer). The remarkable feature of this nonlinear optical process is that all the high-order components are generated collinearly without being restricted by the (angle) phase-matching condition, because the high coherence produced allows us to efficiently generate the high-order Raman components, $E_q$, within a unit phase-slip length [12, 13, 14, 16]. Here, we further introduce another laser field, $E_0^T$, collinearly with the two driving laser fields, $E_0$ and $E_{-1}$. This third laser field is also deeply modulated by the same vibrational motion with high coherence, $\rho_{01}$ (produced above); moreover, it efficiently generates another series of high-order Stokes and anti-Stokes components, $E_q^T$ ($q$: integer), also collinearly without being restricted by the phase-matching condition [14, 17, 18].

This nonlinear optical process can be described by using the Maxwell-Bloch equations [12]. A set of density matrix equations, Eq. 1, expresses the optical Bloch equation for coherent vibrational motion in parahydrogen. In this nonlinear optical process, the medium constitutes a so-called far-off resonant $\Lambda$-scheme [12, 15] and can be effectively reduced to a two-level system by defining the two-photon Rabi frequency $\Omega_{01}$ [12]. We assume that the driving and generated Raman fields propagate in the $z$ direction, and we use the local-time coordinates $\tau = t - z/c$ and $\xi = z$. The constant $c$ is the velocity of light in a vacuum.

$$\frac{\partial \rho_{00}}{\partial \tau} = i(\Omega_{01}\rho_{01}^* - \Omega_{01}^*\rho_{01}) + \gamma_a \rho_{11}$$

$$\frac{\partial \rho_{11}}{\partial \tau} = -i(\Omega_{01}\rho_{01}^* - \Omega_{01}^*\rho_{01}) - \gamma_b \rho_{11} \qquad \text{Eq. 1.}$$

$$\frac{\partial \rho_{01}}{\partial \tau} = i(\Omega_{00} - \Omega_{11} + \delta + i\gamma_c)\rho_{01} + i\Omega_{01}(\rho_{11} - \rho_{00})$$

Here, $\Omega_{00}$ and $\Omega_{11}$ are ac-Stark shifts for the states $|v=0\rangle$ and $|v=1\rangle$, respectively, and $\rho_{00}$, $\rho_{11}$, and $\rho_{01}$ are the population of the ground state $|v=0\rangle$, that of the vibrationally excited state $|v=1\rangle$, and the coherence associated with this Raman transition, respectively. The coefficients $\gamma_a$, $\gamma_b$ and $\gamma_c$, are the decay rates of the populations and of coherence, respectively.

The high-order Raman components, $E_q$, including the two driving-laser fields, $E_0$ and $E_{-1}$, propagate in the nonlinear optical medium according to the Maxwell equation, where all the Raman components are coupled with each other through the Raman coherence, $\rho_{01}$, as below:

$$\frac{\partial E_q}{\partial \xi} = i\frac{N\hbar\omega_q}{\varepsilon_0 c}\left(a_q \rho_{00} E_q + b_q \rho_{11} E_q + d_{q-1}\rho_{01}^* E_{q-1} + d_q^* \rho_{01} E_{q+1}\right) \qquad \text{Eq. 2.}$$

These coupled propagation-equations are expressed in the local-time coordinates with the slowly-varying-envelope-approximation. $E_q$ and $\omega_q$ are the electric field amplitude and angular frequency of the $q$th Raman mode, respectively. $N$, $h$, and $\varepsilon_0$ are the molecular density, Planck constant, and electric permittivity, respectively. The constants $a_q$ and $b_q$ determine the dispersions of the medium and $d_q$ determines the coupling between neighboring Raman components. (See Refs. 12 and 19 for detailed definitions.) The third laser field, $E_0^T$, and its high-order Raman components, $E_q^T$, are also in accordance with the same coupled propagation equation as Eq. 2. Here, as described already, the Raman coherence $\rho_{01}$ is common to the two sets of coupled propagation-equations. However, if we set the third-field amplitude sufficiently weakly compared with those of the two driving fields, then we can treat the behaviors of two series of high-order Raman components, $E_q$ and $E_q^T$, almost independently.

In the above framework, we study the artificial manipulation of high-order Stokes and anti-Stokes generation originating in the third laser field, $E_0^T$, by controlling the relative phase relationship. To see this phase relationship more explicitly, we transform Eq. 2 to Eqs. 3, 4 (*see Supplement*):

$$\frac{\partial \sqrt{n_q}}{\partial \xi} = \frac{N\hbar|\rho_{01}|}{\varepsilon_0 c}\Big\{d_{q-1}\sqrt{\omega_{q-1}\omega_q}\sin(\phi_q - \phi_{q-1} + \phi_\rho)\sqrt{n_{q-1}} - d_q^*\sqrt{\omega_q \omega_{q+1}}\sin(\phi_{q+1} - \phi_q +$$

$$\phi_\rho)\sqrt{n_{q+1}}\Big\}$$

Eq. 3

$$\frac{\partial \phi_q}{\partial \xi} = \frac{N\hbar|\rho_{01}|}{\varepsilon_0 c}\left(a_q\omega_q\frac{\rho_{00}}{|\rho_{01}|} + b_q\omega_q\frac{\rho_{11}}{|\rho_{01}|} + d_{q-1}\sqrt{\omega_{q-1}\omega_q}\cos(\phi_q - \phi_{q-1} + \phi_\rho)\sqrt{\frac{n_{q-1}}{n_q}} + \right.$$
$$\left. d_q^*\sqrt{\omega_q\omega_{q+1}}\cos(\phi_{q+1} - \phi_q + \phi_\rho)\sqrt{\frac{n_{q+1}}{n_q}}\right) \quad \text{Eq. 4.}$$

Here, the electric field at the $q$th order, $E_q$, and the Raman coherence, $\rho_{01}$, are expressed explicitly with the amplitude and phase, as $E_q = |E_q|\exp(i\phi_q)$, $\rho_{01} = |\rho_{01}|\exp(i\phi_\rho)$. Furthermore, we have changed the expression from one regarding field amplitude, $E_q$, to photon number-density per mode, $n_q$. This is done because we want to see this nonlinear optical process from the perspective of "photon flow." Because the total number of photons is conservative in this nonlinear optical process, the set of coupled propagation-equations in Eq. 3 represents a redistribution motion of photon number-densities among all the generated Raman components, including the incident third laser field, $E_0^T$.

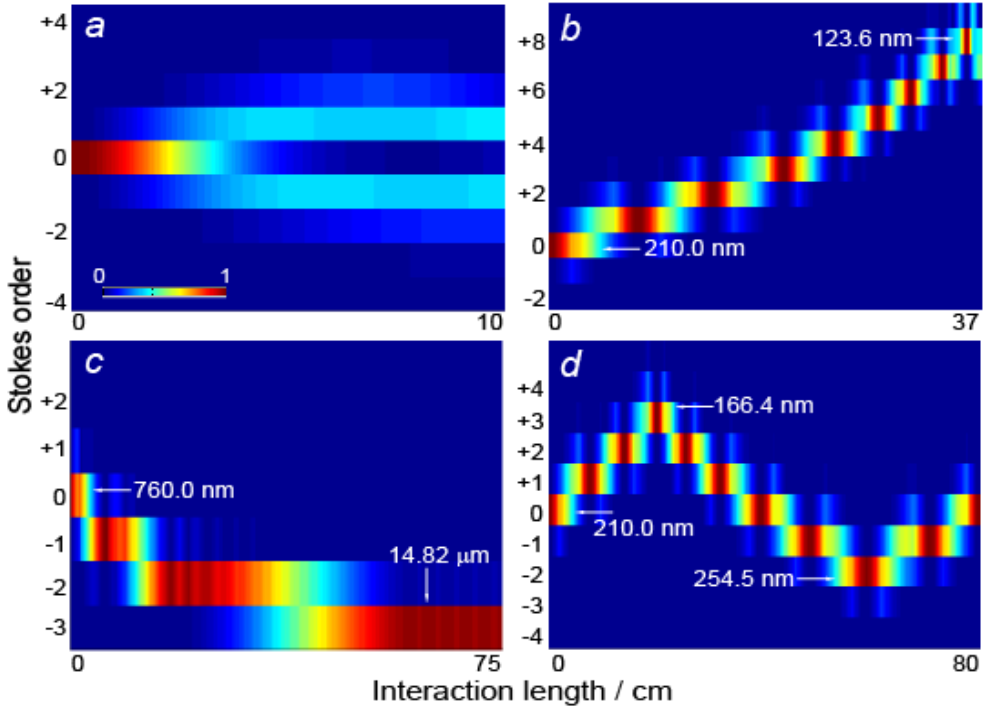

**Fig. 2. Arbitrary manipulation of Raman-resonant four-wave-mixing processes in parahydrogen.** The relative phases among the Raman components, $E_q^T$, are assumed to be controlled arbitrarily.

As seen clearly in this equation, the directions of these photon flows (the first and second terms are the photon flows to the $q$th order from the $(q–1)$th and $(q+1)$th orders, respectively), are determined only by the signs of the relative phases, $\phi(q, q-1) = \phi_q - \phi_{q-1} + \phi_\rho$. Therefore, if we can manipulate such signs, then we can expect to control the "photon flows." As an example, we study the case in which the target of the photon-flow manipulation is set to the photon-number concentration for a specific Raman mode. The following relative phase or sign relationship is typical for realizing such an aim: $mod[\emptyset(q-1, q-2), 2\pi] = \frac{\pi}{2}, mod[\emptyset(q, q-1), 2\pi] = \frac{\pi}{2}$, $mod[\emptyset(q+1, q), 2\pi] = -\frac{\pi}{2}, mod[\emptyset(q+2, q+1), 2\pi] = -\frac{\pi}{2}$. This is because, according to Eq. 3, this relative-phase relationship should give a steep photon flow such that q-2 → q-1 → q ← q+1 ← q+2. We can naturally expect that all the photons distributed among the high-order Raman modes are concentrated to the *q*th order. Furthermore, if we repeat this relative-phase control step by step, we can manipulate the photon flow as intended and thus finally transfer a substantial quantity of the incident photons to the target Raman mode.

To verify this expectation quantitatively, we performed numerical calculations on this nonlinear optical process. Here, we assumed close-to-ideal boundary conditions to confirm the intrinsic potential of the idea as a first step. Namely, we assumed a uniform high Raman coherence of $\rho_{01}$ = 0.3 in both time and space. We also assumed control of the relative-phases to arbitrary values, which were embedded at the optimal interaction lengths in the respective nonlinear optical processes. Under these assumptions, we numerically solved the coupled propagation-equations for the Raman-resonant four-wave mixing process originating in the third laser field, $E_0^T$.

The results in Fig. 2 are typical of those obtained. The molecular density, *N*, and the wavelength of the incident third-laser field were set to $2.6 \times 10^{18}$ $cm^{-3}$ and 210.0000 nm (Figs. 2*a*, *b*, *d*) or $2.6 \times 10^{19}$ $cm^{-3}$ and 760.0000 nm (Fig. 2*c*), respectively. In Fig. 2*a*, as a reference, no phase manipulation was applied. As already described, the incident photons were distributed broadly to the high-order Raman components. In contrast, in Figs. 2*b* to 2*d* various artificial relative-phase manipulations were tested. Such manipulations were performed essentially as indicated above, but, depending on the case we divided the manipulation into more than two steps and thus achieved the desired photon-flows. (*See Supplement for details.*) In Fig. 2*b* the targets were set as the respective high-order Raman modes on the short-wavelength side. As seen clearly, the incident photons were concentrated sequentially down to the 8th order (123.5980 nm), with quantum efficiencies close to unity (> 99%). Here, the relative-phase controls were employed discretely 16 times over the whole interaction length. For every photon-concentration for a specific Raman mode, two steps were required in the relative-phase manipulations. As seen by the clear color-changes, the photons were steeply transferred to the next order around such phase controls. Similarly, we examined the sequential concentrations of photons transferred to the opposite, long-wavelength, side (Fig. 2*c*). The efficiencies were also close to unity (>98%) for all the steps. In this numerical calculation, the longest wavelength was designed to be 14.8203 μm in mid-infrared. The concentration of photons transferred to the long-wavelength side required many more relative-phase manipulation steps, because the ratios of the nonlinear coupling-coefficient to the linear dispersion-coefficient decreased as the relevant laser-wavelength increased. The photon-flow controls are not necessarily restricted to simple unidirectional manipulations like those in Figs. 2*b* and 2*c*. Potentially, arbitrary manipulation is possible. The photon-flow manipulation like a wave in Fig. 2*d* is an example to demonstrate such ability.

There are various ways of realizing the same results by using methods other than this relative-phase manipulation. However, for practical systems, the simpler the control, the better. In this respect, the step-by-step photon-transfer approach described above would be one of the best, because it requires a minimum number of relative-phase controls (typically four).

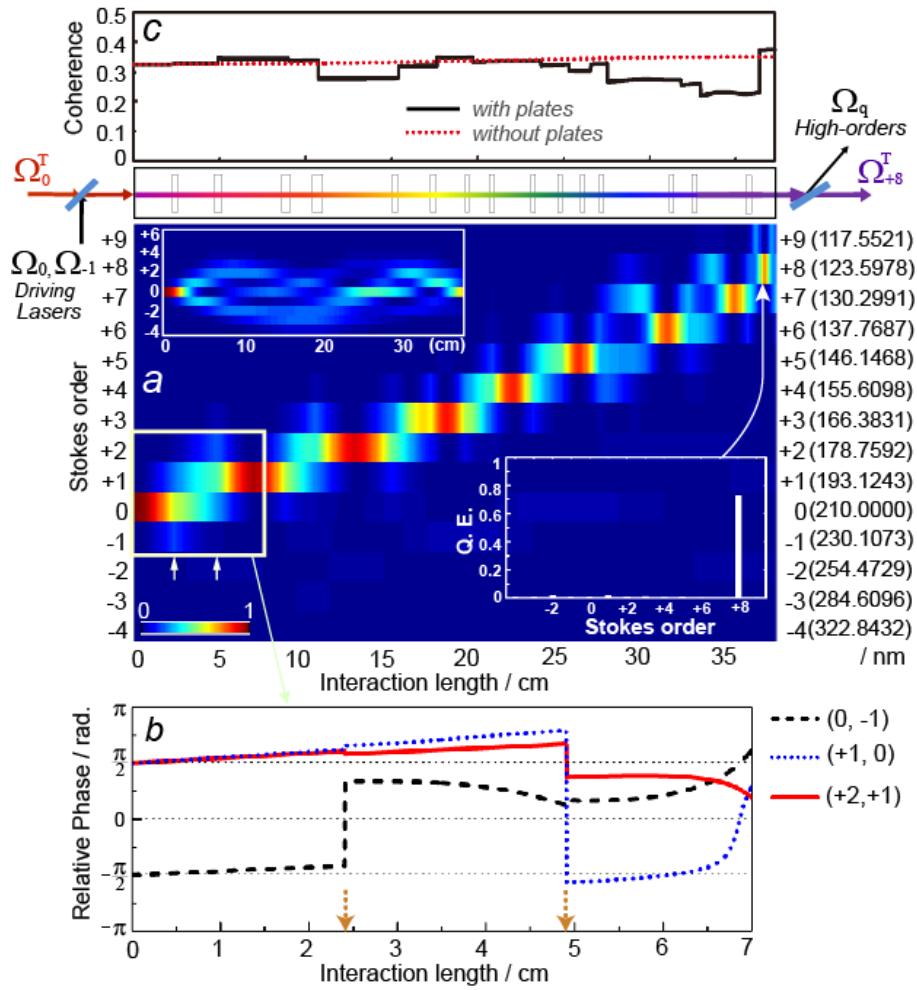

**Fig. 3. Numerical experiment on artificial manipulation of Raman-resonant four-wave-mixing processes in parahydrogen.** *a*, Contour plot of photon-number distributions among high-order Raman modes. *b*, Typical example of relative-phase manipulation by inserting magnesium fluoride plates. *c*, Spatial distribution of vibrational coherence with and without the plates.

Thus, as hoped, we can manipulate photon flows in nonlinear optical processes if we can arbitrarily manipulate the relative phases among the relevant electromagnetic fields. However, how can we achieve such arbitrary phase (or sign) manipulations in reality? Here, we show that they can be realized by using a surprisingly simple technology.

The relative phases among the high-order Raman components can be almost arbitrarily manipulated simply by inserting transparent plates on the optical path and then precisely adjusting their thicknesses (arbitrary optical-phase manipulation by precise control of thickness of a dispersive plate: APM-DiP). This technology can act practically when the frequencies of the relevant electromagnetic fields are discrete, with very large frequency spacings (> tens of terahertz), although it may seem incompatible with the natural physical order. For details, see references 20 to 23; here, we comment only briefly on the key mechanism of the technology. When the frequencies of the relevant electromagnetic fields are very discrete, the thin plates include substantial high-order refractive-index dispersions, and a slight change in their thickness comprehensively sweeps the relative weights among the high-order dispersions. In other words, we can realize almost arbitrary relative-phase relationships.

The critical issue in the actual experiment is that inserting such plates inevitably also affects the process of adiabatic driving of the high coherence, $\rho_{01}$, by the two driving-laser fields, $E_0$ and $E_{-1}$. This coherence-driving process is accompanied by the generation of high-order Raman components, $E_q$. Moreover, it constitutes a physically self-consistent system with such high-order Raman components including their relative phases; this differs completely from the "photon-flow manipulation" in Fig. 2. Therefore, ideally, insertion of the dispersive plates has to simultaneously satisfy these two requirements regarding the controls of the relative-phase relationships: one is for the photon-flow manipulation and the other for adiabatic driving of the high coherence, $\rho_{01}$. However, the coherence itself can be substantially driven also by the two driving-laser fields, alone. Thereby, if we drive the coherence in such regime, the relevant relative-phase can be reduced to only one relative-phase between $E_0$ and $E_{-1}$. This implies that the coherence-driving process is essentially not affected by insertion of the dispersive-plates. Here, influence of inserting the plates still remains as steep changes of the phase of the coherence, $\phi_\rho$, at the plates. Nevertheless, we can manipulate the photon flow as desired, because, as understood from Eq. 3, such changes of $\phi_\rho$ can be included in the relative-phase control for the photon-flow manipulation, $\phi(q, q-1) = \phi_q - \phi_{q-1} + \phi_\rho$.

Our demonstration is also based on a numerical experiment, the numerical code of which has been verified to be highly reliable [12, 13, 14, 19]. Unlike in the former case (Fig. 2), here we treated all the processes realistically by assuming that the experiment was real. The density of gaseous parahydrogen was set to $2.6 \times 10^{18}$ cm$^{-3}$ and the interaction length to 37.2(95) cm. The vibrational coherence, $\rho_{01}$ (124.7451 THz), was adiabatically driven from the ground state by the two-color laser fields ($E_0$: 801.0817 nm; $E_{-1}$: 1201.6261 nm; $\delta = -500$ MHz). The peak intensity of the driving lasers was set at 10 GW/cm$^2$ with a 10-ns pulse duration. The peak intensity of the third laser ($E_0^T$: 210.0000 nm; 5 ns) was set at 0.1 GW/cm$^2$ —100 times weaker than those of the coherence-driving lasers—to avoid interference between the two nonlinear optical processes. The driving and third laser beams were coupled and decoupled in space by setting their polarizations orthogonally (see Fig. 3). As a transparent dispersive material we used magnesium fluoride (MgF$_2$) plates (ordinary axis), because they have high transparency in the vacuum-ultraviolet spectral region. Although not serious, we also took into account absorption losses in the MgF$_2$ plates. For the refractive index dispersion of MgF$_2$ we relied on that given by the Sellmeier equation [24].

To reproduce the optical-frequency conversion demonstrated in Fig. 2*b*, but under actual experimental conditions, we explored the optimum MgF$_2$ plate thicknesses by using the random search method [25], such that they satisfied the requirements for the relative-phase relationship among $E_q^T$. We inserted these plates with the optimum thicknesses at appropriate interaction lengths in gaseous parahydrogen. With this experimental setup, we numerically solved the Maxwell-Bloch equation (Eqs. 1 and 2) for the Raman-resonant four-wave-mixing processes originating in both the driving lasers and the third laser, including their coupling through the coherence, $\rho_{01}$, although the latter is not essential. We did not introduce any artificial assumptions. All of the Raman components were simply retarded in phase to a degree depending on the optical thicknesses of the inserted MgF$_2$ plates when they were transmitted the plates.

Figure 3 gives a typical result. Photon flow very similar to that observed in Fig. 2*a* was reproduced in this numerical experiment. The quantum efficiencies from the incident third laser field to the high-order Raman components were extremely high: 94% at the 1st (193.1243 nm), 84% at the 4th (155.6098 nm) and 73% at the 8th (123.5978 nm). The inset at bottom right of Fig. 3a shows the spectrum generated at an interaction length of 37.2(95) cm. A virtually single line is seen at 123.5978 nm (8th order). To realize this sequential photon-flow manipulation we inserted 15 $MgF_2$ plates. The approximate positions of the plates are illustrated in the panel above Fig. 3a. (S*ee Supplement for details.*) Each plate typically had a position allowance of about ±5 mm and a thickness allowance of about ±1 μm, indicating that the technology was practical. Although 15% of incident photons were lost by absorption onto these 15 $MgF_2$ plates, this was not a serious problem. Figure 3b indicates in detail how we artificially manipulated the relative-phase relationship to realize this photon flow. Only the start of the process is shown (marked with the white square). When the Raman components were transmitted the first $MgF_2$ plate, the relative phase between –1st and 0th was changed such that $0 < \mod[\phi(0,-1), 2\pi] < \pi$. Photon flow to the −1st then flowed back to the 0th as -1 → 0 → +1 → +2. Next, when the Raman components were transmitted the second plate, the relative phase between the +1st and +2nd was controlled as $-\pi < \mod[\phi(+2,+1), 2\pi] < 0$, whereas the others maintained the same sign. The photon flow then changed such that -1 → 0 → +1 ← +2 and finally concentrated at the +1st (193.1243 nm), with 94% concentration at an interaction length of 7.0(8) cm. The inset at top left of Fig. 3a shows the Raman generation with no plates inserted. As in Fig. 2a, the incident photons were naturally distributed broadly to the high-order Raman components. As already described, the above-described phase manipulation by the $MgF_2$ plates must not disturb the coherence-driving process by the two driving-laser fields, $E_0$ and $E_{-1}$. Figure 3c shows the coherence with and without the plates as a function of interaction length at the peak of the driving nanosecond lasers. The 15 inserted $MgF_2$ plates did not spoil a uniform coherence distribution in space, as expected.

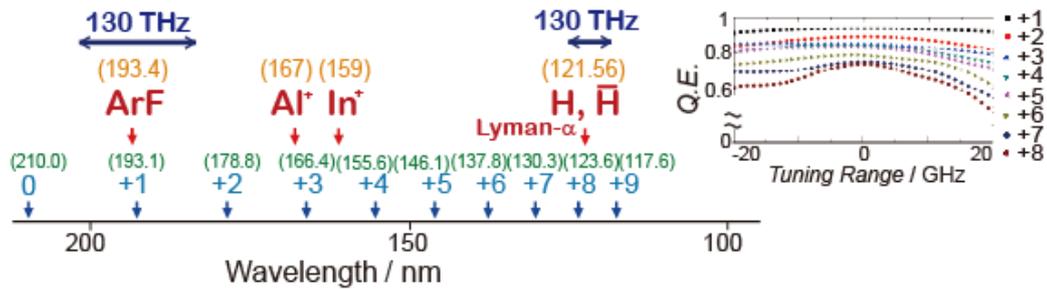

Fig. 4. Tunability of high-order Raman lasers produced by AMNOP, and possible applications of these single-frequency tunable lasers in the vacuum-ultraviolet region of 120 to 200 nm.

Finally, we show some attractive applications of this technology, including precision spectroscopy in the vacuum-ultraviolet region. The third laser field (210.0000 nm) can be practically generated by taking the fourth harmonic of 840.0000 nm which is produced by an injection-seeded

Ti:sapphire laser [26, 27, 28]. Because the tuning range of this laser can be as wide as ±40 nm [27, 28], the 210-nm third laser is tunable over ±10 nm, corresponding to a frequency tuning range of 130 THz—greater than the frequency spacing of the present Raman modes (125 THz). Thereby, we can access any wavelengths from 200 to 120 nm in the vacuum-ultraviolet region. New-wavelength selection requires additional exploration of the optimum thicknesses of the inserted $MgF_2$ plates. However, if the tuning range is within ±20 GHz (sufficient for various spectroscopic applications) we need not adjust the plate thicknesses (see inset which shows quantum efficiencies of the high-order Raman generations where the plate thicknesses were fixed). We also note that, besides this arbitrary wavelength selectivity, this laser technology has other attractive abilities, such as high spectral intensity enabling nonlinear spectroscopy, high frequency precision derived from an optical-frequency-standard precision [29, 30], and scalability to ultrahigh energy (e.g. >1 Joule per pulse).

Some attractive candidates for application are shown in Fig. 4. They include single-ion spectroscopy for optical-frequency standards: $^1S_0$ to $^1P_1$ transition of $Al^+$ (167 nm) [31] and that of $In^+$ (159 nm) [32]; laser cooling of hydrogen and antihydrogen for testing the standard theory: Lyman-$\alpha$ transition at 121.56 nm [33, 34]. Scalability to ultrahigh power will also be attractive from an industry perspective (e.g. high-average-power 193-nm laser for lithography).

In summary, we have described the physics of arbitrary manipulations of nonlinear optical processes. By employing a Raman resonant four-wave mixing process in gaseous parahydrogen, we showed that this nonlinear optical process can be freely controlled by manipulating the relative-phase relationship among the relevant electromagnetic fields. Furthermore, we showed that these arbitrary manipulations can be realized by using an extremely simple technology that inserts transparent dispersive materials ($MgF_2$ plates) into the nonlinear optical medium and adjusts their positions and thicknesses precisely. As a typical example, we demonstrated freely designable optical-frequency conversions with near-unity quantum efficiencies of 73% to 94%; they covered the whole vacuum-ultraviolet spectral range of 120 to 200 nm with a continuous frequency-tunability of more than ±20 GHz.

The concept described in this Letter can be applied to various nonlinear optical processes with various configurations, such as intracavity geometry [35]. Harmonic generations including high-order harmonics, and soliton generation will be attractive candidates. We are currently performing an experiment to verify this concept.

The authors thank K. Ishikawa, K. Yoshii, T. Ido, and F. Shimizu for useful discussions. MK acknowledges the support of a Grant-in-Aid for Scientific Research (A), a Grant-in-Aid for Scientific Research on Innovative Areas, and the Matsuo Foundation.

## References

1. J. A. Armstrong, N. Bloembergen, J. Ducuing, P. S. Pershan, Interactions between light waves in a nonlinear dielectric. Phys. Rev. **127**, 1918-1939 (1962).

2. P. A. Franken, A. E. Hill, C. W. Peters, and G. Weinreich, Generation of optical harmonics. Phys. Rev. Lett. **7**, 118-119 (1961).

3. N. Bloembergen, *Nonlinear Optics* (Benjamin, New York, 1965).

4. Y. R. Shen, *The Principles of Nonlinear Optics* (Wiley-Interscience, New York, 1984).

5. R. Boyd, *Nonlinear Optics* (Academic Press, New York, ed. 3, 2008).

6. J. A. Giordmaine, Mixing of light beams in crystals. Phys. Rev. Lett. **8**, 19–20 (1962).

7. P. D. Maker, R. W. Terhune, N. Nisenoff, and C. M. Savage, Effects of dispersion and focusing on the production of optical harmonics. Phys. Rev. Lett. **8**, 21-22 (1962).

8. N. Bloembergen, The Stimulated Raman Effect. American Journal of Physics **35**, 989-1023 (1967).

9. D. Feng, N. Ming, J. Hong, Y. Yang, J. Zhu, Z. Yang, and Y. Wang, Enhancement of second-harmonic generation in LiNbO3 crystals with periodic laminar ferroelectric domains, Appl. Phys. Lett. **37**, 607-609 (1980).

10. G. A. Magei, M. M. Fejer, and R. L. Byer, Quasi-phase-matched second-harmonic generation of blue light in periodically poled LiNbO3, Appl. Phys. Lett. **56**, 108-110 (1990).

11. H. Suchowski, K. O'Brien, Z. J. Wong, A. Salandrino, X. Yin, and X. Zhang, Phase mismatch-free nonlinear propagation in optical zero-index materials. Science **342**, 1223-1226 (2013); DOI: 10.1126/science. 1244303.

12. S. E. Harris and A. V. Sokolov, Broadband Spectral Comb Generation with Refractive Index Control. Phys. Rev. A **55**, R4019-4022 (1997).

13. A. V. Sokolov, D. R. Walker, D. D. Yavuz, G. Y. Yin, and S. E. Harris, Raman Generation by Phased and Antiphased Molecular States. Phys. Rev. Lett. **85**, 562 - 565 (2000).

14. J. Q. Liang, M. Katsuragawa, Fam Le Kien, and K. Hakuta, Sideband generation using strongly driven Raman coherence in solid hydrogen. Phys. Rev. Lett. **85**, 2474-2478 (2000).

15. K. Hakuta, M. Suzuki, M. Katsuragawa, and J.Z. Li: Self-Induced Phase-Matching in Parametric Anti-Stokes Stimulated Raman Scattering. Phys. Rev. Lett. **79**, 209-212 (1997).

16. M. Katsuragawa, J. Q. Liang, J.Z. Li, M. Suzuki and K. Hakuta: Stimulated Raman Scattering in Solid Hydrogen based on Adiabatic Preparation of Anti-Phased State. CLEO/QELS '99, Technical Digest, QthE2, 195-196, Baltimore, USA, May 23-28 (1999).

17. M. Katsuragawa, J. Q. Liang, Fam Le Kien, and K. Hakuta, Efficient frequency conversion of incoherent fluorescent light. Phys. Rev. A, **65**, 025801-025804 (2002).

18. S. E. Harris, Electromagnetically Induced Transparency. Physics Today **50** (7), 36-42 (1997).

19. Fam Le Kien, J. Q. Liang, M. Katsuragawa, K. Ohtsuki, K. Hakuta, and A. V. Sokolov, Subfemtosecond pulse generation with molecular coherence control in stimulated Raman scattering. Phys. Rev. A **60**, 1562-1571 (1999).

20. K. Yoshii, J. K. Anthony, and M. Katsuragawa, The simplest rout to generating a train of attosecond pulses. Light: Science & Applications, (2013) **2**, e58; DOI:10.1038/lsa.2013.14.

21. K. Yoshii, Y. Nakamura, K. Hagihara, and M. Katsuragawa: Generation of a train of ultrashort pulses by simply inserting transparent plates on the optical path. CLEO/QELS 2014, Technical Digest FThD1.5, San Jose, California, USA, June. 8-13 (2014).

22. T. Suzuki, and M. Katsuragawa, Femtosecond ultrashort pulse generation by addition of positive material dispersion. Optics Express **18**, 23088 – 23094 (2010).

23. Arbitrary manipulation of optical waveforms by simply using three fundamental optical-elements.  *to be submitted*

24. *Handbook of Optics*, 3rd edition, Vol. 4. McGraw-Hill 2009.

25. M. A. Schumer and K. Steiglitz, Adaptive step size random search. IEEE Transactions on Automatic Control **13** (3), 270–276 (1968).

26. C. E. Hamilton, Single-frequency, injection-seeded Ti:sapphire ring laser with high temporal precision. Optics Letters **17**, 728-730 (1992).

27. M. Katsuragawa and T. Onose, Dual-Wavelength Injection-Locked Pulsed Laser. Optics Letters **30**, 2421-2423 (2005).

28. T. Onose and M. Katsuragawa, Dual-wavelength injection-locked, pulsed laser with precisely predictable performance. Optics Express **15**, 1600-1605 (2007).

29. M. Katsuragawa, T. Suzuki, K. Shiraga, M. Arakawa, T. Onose, K. Yokoyama, F. L. Hong, and K. Misawa: Ultrahigh-repetition-rate pulse train with absolute-phase control produced by an adiabatic Raman process. Laser Spectroscopy, Proceedings of the XIX International Conference, ICOLS 2009, ISBN-13 978-981-4282-33-8, World Scientific (2010).

30. T. Suzuki, M. Hirai, and M. Katsuragawa, Octave-spanning Raman comb with carrier envelope offset control. Phys. Rev. Lett. **101**, 243602 - 243605 (2008).

31. P. O. Schmidt, T. Rosenband, C. Langer, W. M. Itano, J. C. Bergquist, and D. J. Wineland, "Spectroscopy using quantum logic," Science **309**, 749-752 (2005).

32. K. Hayasaka and T. Ido, *Private communication.*

33. Continuous Coherent Lyman- *α* Excitation of Atomic Hydrogen, K. S. E. Eikema, J. Walz, and T. W. Hänsch Phys. Rev. Lett. 86, 5679-5682 (2001).

34. The ALPHA Collaboration, "Confinement of antihydrogen for 1,000 seconds," *Nature Physics* **7**, 558–564 (2011).; doi:10.1038/nphys2025.

35. M. Katsuragawa, R. Tanaka, H. Yokota, and T. Matsuzawa, Raman-type optical frequency comb adiabatically generated in an enhancement cavity. arXiv1001.0238v.